\documentclass[onecolumn]{revtex4}
\usepackage{amsmath}

\usepackage{amsfonts}
\usepackage{graphics}
\usepackage{graphicx}
\usepackage{epstopdf}
\usepackage{color}
\usepackage{graphicx}
\usepackage{dcolumn}
\usepackage{bm}
\input epsf
\begin{document}
\title{\Large Wormhole Solutions in Rastall Gravity Theory}

\author{Shibaji Halder}
\email{shibajihalderrrrm@gmail.com}
\affiliation{Department of Mathematics, Vidyasagar College, Kolkata-700006, India}

\author{Subhra Bhattacharya}
\email{subhra.maths@presiuniv.ac.in}
\affiliation{Department of Mathematics, Presidency University, Kolkata-700073, India}

\author{Subenoy Chakraborty}
\email{schakraborty@math.jdvu.ac.in}
\affiliation{Department of Mathematics, Jadavpur University, Kolkata-700032, India}

\begin{abstract}
The present work looks for new wormhole solutions in the non conservative Rastall gravity. Although Rastall gravity is considered to be a higher dimensional gravity, the actual diversion from general relativity essentially happens due to a modification in the corresponding matter tensor part. Thus it would be interesting to find out if such non minimal coupling has any effect on the traversable wormholes and their corresponding energy conditions.
\end{abstract}


\maketitle
\section{Introduction}

In 1972 P. Rastall \cite{rast1} introduced a generalization on the conservation principles of the energy momentum tensors, arguing in favour of $T^{\nu}_{\mu;\nu}\neq 0$ in a curved space-time framework, keeping the original Bianchi identities on the Einstein tensor $G_{\mu\nu}$ unaltered. Using phenomenological justifications he proposed that the divergence of $T^{\mu\nu}$ is such that $T^{\mu}_{\nu;\mu}=\lambda R_{,\nu}$ where $\lambda$ was called the Rastall parameter, $R$ the usual Ricci scalar. This resulted in the corresponding modification of the Einstein tensor,  $G_{\mu\nu}^{(r)}=G_{\mu\nu}+\kappa_{r}\lambda g_{\mu\nu}R$ with $G_{\mu\nu}^{(r)}$ being the modified Einstein tensor and $\kappa_{r}$ a constant. After some rearrangements the above gave the field equations \begin{equation}
G_{\mu\nu}=\kappa_{r} S_{\mu\nu}\label{fldr}
\end{equation} where 
\begin{equation}
S_{\mu\nu}=T_{\mu\nu}-\frac{\kappa_{r}\lambda}{4\kappa_{r}\lambda-1}g_{\mu\nu}T,\label{meff}
\end{equation}
(with $\kappa_{r}\lambda\neq \frac{1}{4}$ hence otherwise $T=0$ would lead to trivial geometries). In the static weak field limit the pure temporal components of the modified equations satisfied 
\begin{equation}
\kappa_{r}\frac{6\kappa_{r}\lambda-1}{8\kappa_{r}\lambda-2} =\frac{\kappa}{2},
\end{equation} where $\kappa$ is the Einstein's gravitational constant. For the weak field scenario of $\lambda=0$ one could obtain $\kappa_{r}=\kappa$ i.e. the usual Einstein's gravity. Further, the above expression showed that the product $\kappa_{r}\lambda=\gamma$ will be a dimensionless parameter and is called the Rastall's dimensionless parameter. For simplicity we assume $\kappa=1,$ then the constant 
\begin{equation}
\kappa_{r}=\frac{4\gamma-1}{6\gamma-1}.\label{kapa}
\end{equation}
Evidently the non-minimal coupling introduced in the matter tensor brought about a corresponding modification in the ensuing gravity description that were consistent with the existing structure of the field equations. Rastall himself suggested the application of his theory to ``cosmology, stellar structures and collapsing objects" \cite{rast1}. Since then multitude of research had been devoted to the cause of applying the Rastall structure for the existing cosmological and astrophysical set up of the universe \cite{rast2,smal,rawa}.

The past decade has seen some renewed vigour to the Rastall cause and new literatures have come up on the various aspects of application of Rastall gravity. Dark sector unification became one of the areas of application and showed promising results. In fact two fluid models in Rastall gravity compared successfully with the $\Lambda$CDM model \cite{fav,bat1,bat2}. Analysis of Abelian-Higgs string in Rastall gravity produced quantum effects that could influence inflationary predictions \cite{mell}. Studies on astrophysical properties of the neutron star under the Rastall gravity framework has established successful bounds on the Rastall parameter \cite{oliv}. About a couple of years ago a thermodynamic analysis of spherically symmetric solutions in Rastall gravity using both first and second laws of thermodynamics had been performed \cite{morad1, morad2}. The past year has seen a surge of activities on description of Black holes and their thermodynamics in the Rastall framework \cite{bh1,bh2,bh3,bh4,bh5}.

Traversable wormholes \cite{1} are astrophysical objects that are described as theoretical means for interstellar travel. The basic recipe for the traversability of such wormholes is the ``no horizon" condition of the throat (passage) such that the throat has minimal surface area and satisfies a flare-out condition \cite{1, 2, 3, 7}. This in turn leads to the violation of null energy condition (NEC) by the matter stress energy tensor at the throat \cite{1,3,7,8}. Exotic matter was proposed for the existence of such wormholes. Several researches have tried to address the nature of the exotic matter under different settings \cite{3, 10, 11, 12}. Modified gravity theories has been one such framework, where the extra geometrical terms are theorized to be the source of such exotic energies \cite{dotti1, dotti2, dotti3, matulich, harko, lobo, la,  sb}. Rastall gravity, with its myriads of applications in astrophysics has emerged as one possible candidate for testing its influence the existence of wormholes. Especially so, in Rastall gravity the actual generalization is taking place w.r.t the energy momentum tensor only, with the corresponding geometric part remaining unaltered. This makes not just the field equations easier to handle, but also raises interesting possibilities on its influence in defining throat matter. Very recently Rastall gravity framework had been used to address asymptotically flat traversable wormholes corresponding to a specific shape function of the throat \cite{morad3}. Our aim in this work would be to find out a new class of viable traversable wormhole solutions in Rastall gravity theory and hence determine the viable conditions for their existence based on appropriate parameter restrictions.

The paper shall be organised as follows: In section 2 we shall explore the possibilities of traversable wormhole solutions in Rastall gravity together with their suitable energy equations. In section 3 we give specific examples of wormholes, both with zero and non zero tidal force type, which can exit in Rastall theory. Section 4 shall end with a brief discussion and conclusion of the results presented.

\section{Traversable Lorentzian Wormholes in Rastall Gravity}
         
In spherically symmetric space-time coordinates the traversable wormhole is usually described by the line element \cite{1}
\begin{equation}
ds^{2} = - e^{2\phi(r)} dt^{2} + \frac{dr^{2}}{1 - \frac{b(r)}{r}}
+ r^{2} d\Omega_{2}^{2}
\end{equation}
where $\phi(r)$ is the gravitational red shift function and $b(r)$ is the shape function indicative of the shape of the wormhole throat. Here, in order for the wormhole to be traversable, formation of an event horizon should be avoided, i.e. $\phi(r)$ should be finite everywhere. Again by definition of a wormhole the shape function should be such that, at the wormhole throat the radial coordinate should be a minima. Then using the proper radial distance $l$ and an embedding surface on the geometry it can be shown that $\frac{dr}{dl}= 0$ at $r=r_{min}=r_{0}$ the wormhole throat. This gives $b(r) = r_{0}$ at the throat. Also for the wormhole to act as a connection between two asymptotically flat space-time, it is required that the throat flare outward. This is given by $\left[b- (b'r)\right]/b^{2} > 0$ at $r= r_{0}$. The prescribed constraint essentially means defocusing of null geodesic congruences at the throat and hence matter at the throat must violate NEC.

The field equations in Rastall gravity is given by (\ref{fldr}). Where $S_{\mu\nu}$ is the effective energy momentum tensor in Rastall gravity (henceforth referred as the Rastall matter) defined by (\ref{meff}). If $(\xi,\sigma_{r},\sigma_{t})$ be the matter stress energy components of $S_{\mu\nu},$ then from (\ref{fldr}) we get
\begin{equation}
\left.\begin{aligned}
\kappa_{r}\xi=&\frac{b'(r)}{r^{2}}\\
\kappa_{r} \sigma_{r}=&-\frac{b}{r^{3}}+2\left(1-\frac{b}{r}\right)\frac{\phi'}{r}\\
\kappa_{r} \sigma_{t}=&\left(1-\frac{b}{r}\right)\left[\phi''-\frac{b'r-b}{2r(r-b)}\left(\frac{1}{r}+\phi'\right)+(\phi')^{2}+\frac{\phi^{'}}{r}\right].
\end{aligned}
\right\}\label{s}
\end{equation}
If $(\rho,p_{r},p_{t})$ the corresponding components for $T_{\mu\nu}$ (henceforth addressed as the actual matter), then using (\ref{meff}) it can be shown that
\begin{equation}
\left.\begin{aligned}
\rho&=(1-\gamma)\xi+\gamma \sigma_{r}+2\gamma \sigma_{t}\\
p_{r}&=\gamma\xi+(1-\gamma)\sigma_{r}-2\gamma \sigma_{t}\\
p_{t}&=\gamma\xi-\gamma \sigma_{r}+(1-2\gamma) \sigma_{t},
\end{aligned}
\right\}\label{t}
\end{equation}
where $\gamma$ is the dimensionless Rastall parameter that has been defined earlier. Using (\ref{s}) and (\ref{t}), for specific shape functions one can evaluate $S_{\mu\nu}$ and hence $T_{\mu\nu}$ to determine the parameter requirements for traversable wormhole. 

One can observe that $p_{r}=p_{t}\Leftrightarrow\sigma_{r}=\sigma_{t}.$ If the isotropic equation of states are defined as $\omega=\frac{p}{\rho}$ and $\eta= \frac{\sigma}{\xi}$ with $p$ and $\sigma$ being the corresponding isotropic pressures, then using (\ref{t}) we can find
\begin{equation}
\eta+1=-\frac{(4\gamma-1)(1+\omega)}{1-3\gamma(1+\omega)}.
\end{equation} 
Thus $\eta=-1\Rightarrow\omega=-1$ and vice-versa ($\gamma\neq \frac{1}{4}$). Also, 
$$\eta+1>0\Rightarrow\omega+1>0~\text{for}~\gamma>max(\frac{1}{4},\frac{1}{3(1+\omega)})~\text{or}~\gamma<min(\frac{1}{4},\frac{1}{3(1+\omega)}).$$
If $$min(\frac{1}{4},\frac{1}{3(1+\omega)})<\gamma<max(\frac{1}{4},\frac{1}{3(1+\omega)})~\text{then}~(1+\eta)< 0~(>0)\Rightarrow (1+\omega)>0~(<0).$$ 

In case of anisotropic matter it can be verified $\xi+\sigma_{r}+2\sigma_{t}=0$ that does not essentially give $\rho+p_{r}+2p_{t}=0.$ However $\rho+p_{r}=\xi+\sigma_{r}$ and $\rho+p_{t}=\xi+\sigma_{t}$ is always fulfilled. Further if $$\rho+p_{r}+2p_{t}\geq 0\Rightarrow\xi+\sigma_{r}+2\sigma_{t}>0~\text{provided}~\gamma\geq \frac{1}{2}$$ and $$\rho+p_{r}+2p_{t}< 0\Rightarrow\xi+\sigma_{r}+2\sigma_{t}\geq 0~\text{provided}~\frac{1}{4}<\gamma<\frac{1}{2}.$$ Thus non-conservation of actual matter does not essentially mean that Rastall matter will be non-conserved. Further one can observe that Rastall matter satisfying strong energy conditions (SEC) does not necessarily mean the same for actual matter.

\section{Specific Wormhole Solutions}

Here we will provide example of wormholes that can exist in Rastall gravity framework. We shall consider both the zero tidal force traversable kind, i.e. $\phi(r)=0$ and ones with non-zero tidal force or general red-shift function. Corresponding to specific shape function we shall establish their existence using their known properties as discussed above. Further we shall also find restrictions on the Rastall parameter such that such wormholes can exist without possible NEC violation.

\subsection{Wormholes with non-zero tidal force}

We shall provide example of two such wormhole with non-zero tidal force or a general red-shift function.

\begin{enumerate}

\item We consider the red shift function as 
\begin{equation}
e^{\phi}=\phi_{1}+\frac{\phi_{0}}{2\alpha}r^{2\alpha}\label{phi1}
\end{equation} and corresponding shape function as 
\begin{equation}
\frac{b(r)}{r}=\left(\frac{r}{r_{0}}\right)^{2\alpha}\left(\frac{\frac{\phi_{0}}{2\alpha\phi_{1}}(2\alpha+1)r_{0}^{2\alpha}+1}{\frac{\phi_{0}}{2\alpha\phi_{1}}(2\alpha+1)r^{2\alpha}+1}\right)^{\frac{2\alpha}{2\alpha+1}}\label{b1}
\end{equation}
with $\alpha<0,~\phi_{0,1}$ being arbitrary constants \cite{us}. Then considering the Einstein's field equations we find that at the throat $r=r_{0}$
\begin{equation}
\kappa_{r}\xi=\frac{1+2\alpha c}{r_{0}^{2}},~\kappa_{r}\sigma_{r}=-\frac{1}{r_{0}^{2}},\label{xi1}
\end{equation}
where $c=\frac{r_{0}e^{\phi(r_{0})}}{\phi_{0}r_{0}^{2\alpha}+e^{\phi(r_{0})}}$ is a positive number.
\begin{description}
\item $\kappa_{r}>0~(\gamma\in(-\infty,1/6)\cup(1/4,\infty))$: Here $\xi>0$ for $-\frac{1}{2c}<\alpha<0.$ Evidently NEC is violated at the throat.  Also $\rho=(1-\gamma)\xi+\gamma(1+2\alpha)\sigma_{r}$ and $p_{r}=\gamma\xi+(1-\gamma-2\alpha\gamma)\sigma_{r}.$ Thus the radial equation of state parameter is given by $$\omega_{r}=\frac{p_{r}}{\rho}=-1+\beta,~\text{where}~\beta=\frac{\xi+\sigma_{r}}{(1-\gamma)\xi+(1+2\alpha)\gamma\sigma_{r}}.$$ It is shown that at the throat $\xi+\sigma_{r}<0.$ Also $(1-\gamma)\xi+(1+2\alpha)\gamma\sigma_{r}=\rho\kappa_{r}r_{0}^{2}>0$ at the throat. Thus $\beta<0\Rightarrow\omega_{r}<-1,$ signifying actual matter make up with phantom characteristics. In fact one can observe that for this case, when $\kappa_{r}=1,$ the Rastall dimensionless constant $\gamma$ and hence the Rastall parameter $\lambda$ vanishes, and the wormhole solution given by (\ref{phi1}) and (\ref{b1}) reduces to that obtained in Einstein gravity \cite{us}. 

\item $\kappa_{r}<0~(1/6<\gamma<1/4)$: Here again $\xi>0$ at the throat provided $\alpha<-\frac{1}{2c}.$ Because $\kappa_{r}<0$ evidently, NEC is not violated at the throat. Further SEC (strong energy conditions) are obeyed for $c>1.$ For this case $\beta>0.$ This gives $-1<\omega_{r}<0$ for $0<\beta<1$ and $0<w_{r}<1$ for $1<\beta<2.$
\end{description}
Thus using our first example itself, we are able to show that general wormhole solutions in Rastall gravity can exist with both normal and phantom matter depending on the behaviour of the Rastall coupling factor. We shall now move on to show some more similar solutions.

\item Our next example has red-shift function specified by 
\begin{equation}
e^{\phi(r)}=\phi_{1}+\phi_{0}\sqrt{1-\left(\frac{r}{r_{0}}\right)^{2\alpha}}\label{phi2}
\end{equation}
 and shape function given by 
\begin{equation}
 \frac{b(r)}{r}=\left(\frac{r}{r_{0}}\right)^{2\alpha}\label{b2}
\end{equation}
with $\alpha<0,~\phi_{0,1}$ all arbitrary constants \cite{us}. Computing the components of Rastall matter we get:
\begin{equation}
\kappa_{r}\xi=\frac{1+2\alpha}{r_{0}^{2}},~\kappa_{r}\sigma_{r}=-\frac{1}{r_{0}^{2}},\label{xi2} 
\end{equation}
all computed at the throat $r_{0}.$ 
\begin{description}
\item $\kappa_{r}>0~(\gamma\in(-\infty,1/6)\cup(1/4,\infty))$: Here $\xi>0$ for $-\frac{1}{2}<\alpha<0.$ NEC is violated at the throat.  At the throat the radial equation of state parameter corresponding to actual matter $$\omega_{r}=\frac{p_{r}}{\rho}=-1+\alpha_{1}\gamma_{1}$$
where $$\alpha_{1}=\frac{2\alpha}{(1+2\alpha)}<0~\text{and}~\gamma_{1}=\frac{3-2\gamma}{1-2\gamma}>0~\text{for}~\gamma\in(-\infty,1/6)\cup(1/4,1/2)\cup(3/2,\infty).$$  
Thus $\alpha_{1}\gamma_{1}<0$ which gives $\omega_{r}<-1$ that is super quintessence/phantom fluid at the throat. For $\gamma\in(1/2,3/2)$ we get $\gamma_{1}<0$ and hence obtain $-1<\omega_{r}<0$ for $0<\alpha_{1}\gamma_{1}<1$ and $0<\omega_{r}<1$ for $1<\alpha_{1}\gamma_{1}<2.$ (It is to be noted that in this example we require $\gamma\neq 1/2$). At $\gamma=0,~\gamma_{1}=3$ and actual matter is same as Rastall matter, also the wormhole solution obtained from (\ref{phi2}) and (\ref{b2}) reduces that in usual gravity \cite{us}, with matter at the throat violating NEC.

\item $\kappa_{r}<0~(1/6<\gamma<1/4)$: Again $\xi>0$ at the throat provided $\alpha<-\frac{1}{2}.$ NEC are always satisfied for the above values of $\kappa_{r}$ and $\alpha.$ Here for the given range of $\alpha$ and $\gamma$ we always get $\alpha_{1}\gamma_{1}>0.$ Thus actual matter can never be in the phantom range. It can be quintessence at most.
\end{description}

\end{enumerate}

\subsection{Wormholes with zero tidal force}

Here we shall provide four separate wormhole examples. Since these wormholes will be characterized by zero-tidal force, these are essentially traversable, i.e. one can use them for the purpose of constructing suitable time-machines \cite{2}. All these wormholes are characterized by anisotropic Rastall and actual matter.

\begin{enumerate}
\item Let us consider the function 
\begin{equation}
b(r)=\frac{r_{0}^{n}}{r^{n-1}}\label{b3}
\end{equation}
for some  $n>0.$ Computing the components of Rastall matter we get:
\begin{equation}
\kappa_{r}\xi=\frac{1-n}{r_{0}^{2}},~\kappa_{r}\sigma_{r}=-\frac{1}{r_{0}^{2}},\label{xi3} 
\end{equation} at the throat $r_{0}.$

\begin{description}
\item $\kappa_{r}>0~(\gamma\in(-\infty,1/6)\cup(1/4,\infty))$: For this we have $\xi>0$ corresponding to $0<n<1.$ Further one can easily observe that NEC is always violated at the throat. Evaluating the components of actual matter we find that $\rho=(1-2\gamma)\xi>0$ for 
$\gamma\in(-\infty,1/6)\cup(1/4,1/2).$ Further the radial equation of state parameter at the throat is given by $\omega_{r}=-1-\frac{n}{(1-2\gamma)(1-n)}.$ This shows that equation of state for actual matter is always in the phantom range for the above range of $\gamma.$ Specifically for vanishing Rastall parameter $\lambda,~\gamma=0$ and we get wormhole solution given by (\ref{b3}) matching to that in usual gravity theory with actual matter and Rastall matter coinciding.

\item $\kappa_{r}<0~(1/6<\gamma<1/4)$: Here $\xi>0$ for $n>1.$ Energy conditions are easily satisfied in the given parameter range. For this range of $n,~n_{1}=\frac{n}{(1-2\gamma)(1-n)}<0.$ This gives quintessence type matter i. e. $-1<\omega_{r}<0$ for $-1<n_{1}<0,$ and for $-2<n_{1}<-1$ one can obtain matter with $0<\omega_{r}<1.$ 
\end{description}

\item We next consider the function 
\begin{equation}
b(r)=\frac{r}{1+(r-r_{0})}\label{b4}
\end{equation}
as the corresponding shape function for $0<r_{0}<1$ \cite{us}. Evaluating the components of Rastall matter at the throat we find that:
\begin{equation}
\kappa_{r}\xi=\frac{1-r_{0}}{r_{0}^{2}},~\kappa_{r}\sigma_{r}=-\frac{1}{r_{0}^{2}},\label{xi4} 
\end{equation} 
Depending upon the choice of the parameter $\kappa_{r}$ we again try to determine valid ranges of the throat $r_{0}$ for which effective matter (Rastall) satisfies the NEC. 

\begin{description}
\item $\kappa_{r}>0~(\gamma\in(-\infty,1/6)\cup(1/4,\infty))$: $\xi>0$ for $0<r_{0}<1.$ NEC is always violated. Further evaluating the components of actual matter we find that $\rho=(1-2\gamma)\xi>0$ for $\gamma\in(-\infty,1/6)\cup(1/4,1/2).$ For this range of $\gamma$ and $r_{0}$ we find that the effective radial equation of state given by $\omega_{r}=-1-\frac{r_{0}}{(1-2\gamma)(1-r_{0})}$ is always in the phantom range. Here again $\gamma=0$ gives $\rho=\xi$ and wormhole result given by (\ref{b4}) coinciding with that in GR and matter make up at the throat violating NEC.

\item $\kappa_{r}<0~(1/6<\gamma<1/4)$: In this scenario $\xi>0$ only when $r_{0}>1.$ This however means that we no longer get a traversable wormhole throat connecting two asymptotically flat regions of space-time, instead we get a wormhole at spatial infinity. For traversable wormholes this value of $\kappa_{r}$ we require effective Rastall matter to have negative energy density. Since this is not very feasible we do not consider this case any further.
\end{description}
This is an example which shows that Rastall matter background may not be always effective for wormholes to exist with no energy condition violation.

\item As a third choice we take $b(r)=\frac{r}{1+\log(1+r-r_{0})}.$ This will be a valid wormhole throat provided $0<r_{0}<1.$ For this $b(r)$ the corresponding values for Rastall energy momentum tensors at $r_{0}$ is given by:
\begin{equation}
\kappa_{r}\xi=\frac{1-r_{0}}{r_{0}^{2}},~\kappa_{r}\sigma_{r}=-\frac{1}{r_{0}^{2}},\label{xi5} 
\end{equation} 

This shows that the matter properties at the throat in this case to similar to example above. Hence for this type of shape function we shall obtain phantom like actual matter for $\kappa_{r}>0,$ while for $\kappa_{r}<0$ the wormhole exists only at spatial infinity, with supporting effective matter satisfying NEC. The resulting radial equation of state for actual matter can be in the quintessence range or radiation type.

\item Here we choose 
\begin{equation}
b(r)=re^{\frac{2\alpha}{\beta}(r^{\beta}-r_{0}^{\beta})}\label{b5}
\end{equation}
for some constant $\alpha<0$ and $\beta>0,$ with $r_{0}$ being the location of the wormhole throat \cite{us}. The components of Rastall parameter being
 \begin{equation}
\kappa_{r}\xi=\frac{1+2\alpha r_{0}^{\beta}}{r_{0}^{2}},~\kappa_{r}\sigma_{r}=-\frac{1}{r_{0}^{2}},\label{xi5} 
\end{equation}
\begin{description}
\item $\kappa_{r}>0~(\gamma\in(-\infty,1/6)\cup(1/4,\infty))$: $\xi>0$ for $-\frac{1}{2r_{0}^{\beta}}<\alpha<0.$ NEC is violated at the throat. Further evaluating the components of actual matter we find that $\rho>0$ for $\gamma\in(-\infty,1/6)\cup(1/4,1/2).$ For this range of $\gamma$ and $\alpha$ we find that the effective radial equation of state for actual matter given by $\omega_{r}=-1+\frac{2\alpha r_{0}^{\beta}}{(1-2\gamma)(1+2\alpha r_{0}^{\beta})},$ that is equation of state of matter is always in phantom range. It is evident from above that vanishing $\gamma$ implies wormhole solution (\ref{b5}) in usual Einstein gravity. 

\item $\kappa_{r}<0~(1/6<\gamma<1/4)$: In this scenario $\xi>0$ only when $-\frac{1}{2r_{0}^{\beta}}>\alpha.$ Here again SEC are satisfied. Further one can see that $-1<\omega_{r}<0$ for $0<\frac{2\alpha r_{0}^{\beta}}{(1-2\gamma)(1+2\alpha r_{0}^{\beta})}<1.$ While $0<\omega_{r}<1$ for $1<\frac{2\alpha r_{0}^{\beta}}{(1-2\gamma)(1+2\alpha r_{0}^{\beta})}<2.$
\end{description}
\end{enumerate}

Thus we could show examples of wormholes that can exist in Rastall gravity with matter satisfying as well as violating NEC. The existence of such wormholes is solely determined by suitable choice of the dimensionless parameter $\gamma.$ Further all these above solutions embodies the usual general relativistic solution of wormhole as a special case of $\gamma=0.$  

\section{Brief Discussion and Conclusion}

In the above work we have obtained new viable wormhole solutions in the Rastall gravity framework. Corresponding to the solutions obtained we have shown that depending on the Rastall parameter one can obtain conditions where traversable wormholes can exist even with effective matter satisfying SEC. Thus the Rastall gravity framework is capable of modifying the energy condition requirements of the matter stress energy energy tensor at the throat so that the matter satisfies SEC. The only fall out of such a requirement is that either the Rastall coupling constant $\kappa_{r}$ or $\lambda$ has to be negative. A negative non minimal coupling takes care of the geometric curvature requirement of the wormhole throat and induces suitable factors for focusing of null geodesic congruences at the wormhole throat such that NEC is not violated. 

However if one considers a positive non minimal coupling then the resulting matter stress energy tensor at throat is most likely to be a phantom fluid. Further in the GR limit of $\lambda\rightarrow 0$ we can always obtain the wormhole solution in Einstein gravity.

We also observe that the nature of the energy conditions of effective energy-momentum tensor and actual ones are always coupled, in the sense that if one satisfies energy conditions the other will also and vice-versa. This is particularly important in case of wormholes, for modification in actual matter results in Rastall matter, such that matter satisfying SEC can support traversable wormholes based on the choice of the modification introduced, whose action is due to a proper choice of Rastall parameter.

Thus the paper provides hosts of examples of such wormholes with valid parameter range that can exist with SEC, in the presence of matter modification introduced with suitable parameter labelling. Indeed Rastall gravity can serve to ease some of problems that plagues wormholes in Einstein gravity.

\section{Acknowledgments}
 SB acknowledges UGC's Faculty Recharge Programme and Department of Science and Technology, SERB, India for financial help through ECR project (File No. ECR/2017/000569). SC thanks IUCAA, Pune, India, for their warm hospitality while working on this project.


\end{document}